\documentclass{aa}
 \usepackage{graphicx}
 \usepackage{natbib}
 \begin{document}

\def\lapp{\ifmmode\stackrel{<}{_{\sim}}\else$\stackrel{<}{_{\sim}}$\fi}
\def\gapp{\ifmmode\stackrel{>}{_{\sim}}\else$\stackrel{>}{_{\sim}}$\fi}

 \title{Structure of pulsar beams: conal versus patchy}
 \author{Jaros\l aw Kijak \and Janusz Gil}

 \offprints{J. Kijak}

 \institute{Institute of Astronomy, University of Zielona G\'ora,
 Lubuska 2, 65-265, Zielona G\'ora, Poland} 

\date{Received 19 November 2001 / Accepted 30 May 2002}

\abstract{
Structure of mean pulsar radiation patterns is discussed within
the nested-cones and patchy beam models. Observational predictions
of both these models are analyzed and compared with available data
on pulsar waveforms. It is argued that observational properties of
pulsar waveforms are highly consistent with the nested-cone model
and, in general, inconsistent with the patchy beam model.
}

\titlerunning{Structure of pulsar beams: conal versus patchy}
\authorrunning{J. Kijak \& J. Gil}

\maketitle

\section{Introduction}

  One of the important questions in pulsar research is what the
 overall structure of the mean pulsar beam is and how this
 structure is related to highly fluctuating instantaneous pulsar
 radiation. It is difficult to reveal this structure as pulsar
 observations represent one-dimensional cuts through
 two-dimensional beams. However, some indirect methods have been
 applied in an attempt to resolve this problem and two major models
 of pulsar beams have emerged from this work. \citet{r93},
 \citet{gks93} and \citet{kwj94} calculated the opening angles
 $\rho$ of emission corresponding to a pulse width $W$ measured at
 10 and 50 percent of the maximum intensity. As a result, they
 obtained a binomial distribution of these angles, that is, for a
 given period $P$ one of the two preferred values was possible,
 following however a general $P^{-1/2}$ dependence. Such 
 distribution is most naturally interpreted as an indication of two
 nested cones in the structure of mean pulsar beams. This
 interpretation is called a conal model of pulsar beams. An
 alternative model postulates that the mean pulsar beam is patchy 
\citep[][LM88 hreafter]{lm88},
 with different components randomly distributed within an almost
 circular ``window function'' \citep{m95,hm01}. Such a model is apparently
 inconsistent with the binomial distribution of the opening angles
 inferred from measured pulse widths. In fact, unless putative patches are
 distributed along nested-circular patterns, the distribution of 
 corresponding opening angles should be (for any given period)
 random rather than binomial.

\begin{figure*}
\centering
\includegraphics[height=18cm]{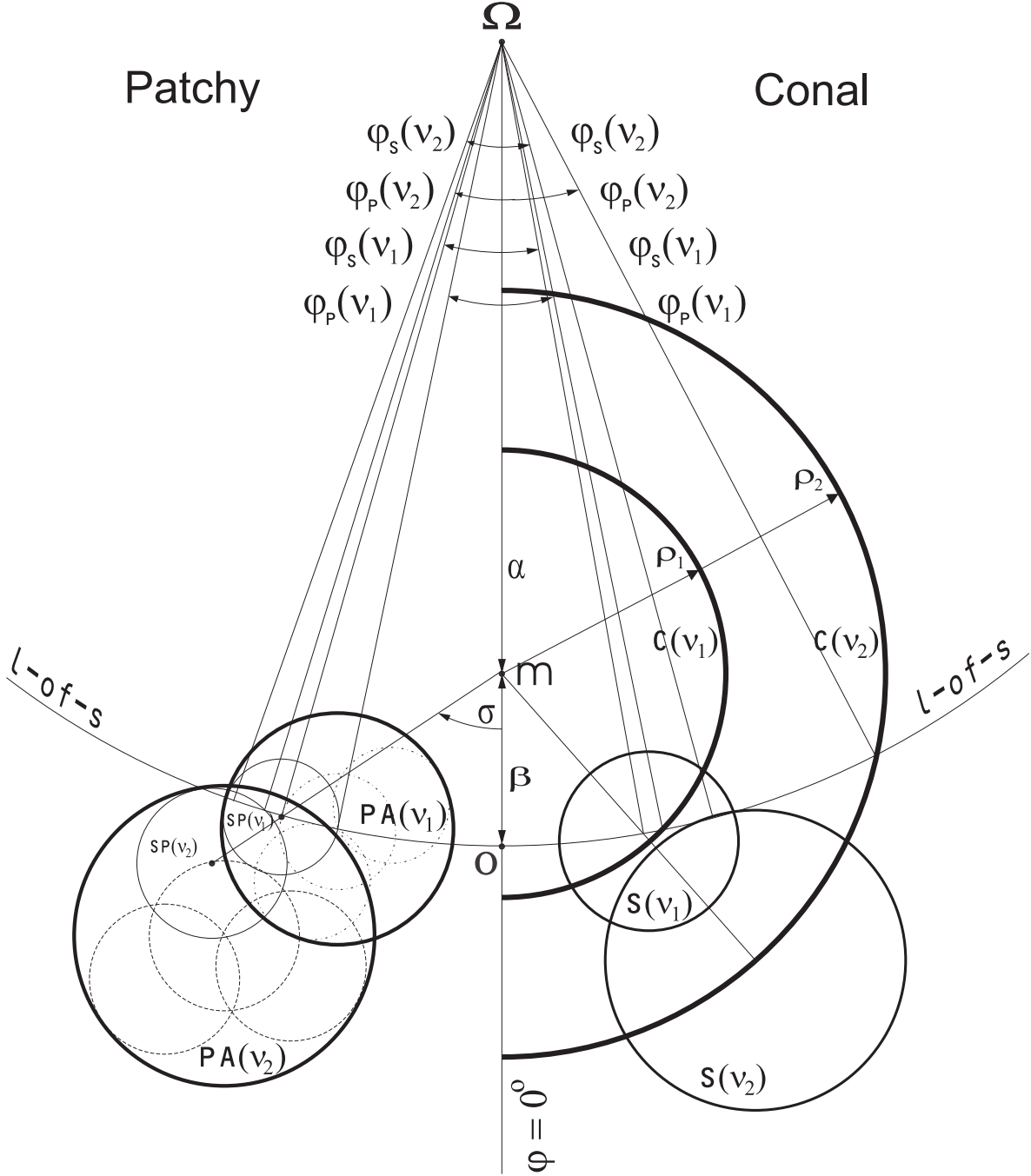}
\caption{The beam emission patterns for both the conal model
(illustrated on the right-hand side), and for the patchy beam
model (illustrated on the left-hand side) are presented in this
composite picture. The geometry of observation is determined in
both models by the inclination angle $\alpha$ between the magnetic
$\bf{m}$ and the spin $\bf\Omega$ axes and the impact angle
$\beta$ of the closest approach of the line-of-sight (observer
${\bf O}$) to the magnetic axis $\bf{m}$. The fiducial
plane containing the
spin axis $\bf\Omega$, the magnetic axis $\bf{m}$ and the observer
$\bf{O}$ defines the fiducial longitude $\varphi=0^\circ$ 
which is common for both models. All pulse phases
($\varphi_s$ - longitude of subpulse peak, $\varphi_p$ - longitude of
profile component peak) are measured from this
fiducial longitude, to the left from it in the patchy model and to
the right from it in the conal model. In this paper we ignore the
dispersive delays which are the only cause of frequency variation at 
or near the fiducial plane. The frequency-dependent
position of any observed feature in the beam pattern is described
by two angles: the opening angle $\rho(\nu)$ between the ${\bf m}$
axis and the line-of-sight (l-of-s) and by the azimuthal angle $\sigma$
between the fiducial plane and the plane of dipolar field lines
associated with a particular feature. Within the conal model
the subpulse emission is
associated with the frequency dependent subpulse spots $S(\nu)$,
which move circumpherentially to form a cone $C(\nu)$.
Thus, the profile components are associated with
the frequency dependent cones in the conal model,
while within the patchy model the
subpulse-associated spots $SP(\nu)$ occupy the patchy areas
$PA(\nu)$ related to the  profile components. The longitudes of
subpulse peaks $\varphi_s(\nu)$ and profile peaks $\varphi_p(\nu)$
are marked, with $\nu_1>\nu_2$, in both models.
\label{fig1}}
\end{figure*}

 Recently, \citet[][MD99 hereafter]{md99} attempted to test both these rival
 models. They distributed locations of the profile components
 (measured as the peak-to-peak separation of the outer conal components
 in complex profile pulsars) on one quadrant of the beam
 represented on the common normalised scale (with $x$-axis and
 $y$-axis representing longitudes $\varphi$ and impact angles
 $\beta$, respectively; refer to Fig.~4 of MD99).
 They found that most of the peak intensity points are
 concentrated in narrow sections of the beam. This feature is a
 strong indication of the conal structure of the pulsar beam. It can
 be argued that if, indeed, the pulsar beams are patchy, then in
 such a case there is a high probability for the beam to be
 uniformly filled with peak intensity locations. Further in their
 analysis, they excluded the so-called conal single and conal
 double profiles \citep{r83}, which are thought to be exclusively
 grazing cuts of the line-of-sight at the beam boundary. They found
 that such exclusion in their sample led to the absence of
 points at high impact angles $\beta$, which is perfectly
 consistent with the nested cone model and inconsistent with the
 patchy beam model. In fact, within the patchy beam model there is
 no reason why the single and double profiles occur exclusively at
 high impact angles. Moreover, the midpoint of single and double
 profiles usually coincides with the fiducial phase, at which the
 multifrequency profiles align after being corrected for cold plasma
 dispersive delays. This property is natural within the conal model
 and inconsistent (in general) with the patchy beam model, since
 patches would have to be placed symmetrically with respect to the
 fiducial plane, containing both magnetic ${\bf m}$ and spin
 ${\bf\Omega}$ axes (Fig.~\ref{fig1}). As argued by MD
 and independently by \citet[][GS00 hereafter]{gs00}
the ``mean'' average pulsar beam
 consists of up to three nested cones, centered on the global
 magnetic dipole axis.

 Observations of single pulses in strong pulsars show that
 longitudes of subpulses are weakly dependent on frequency as
 compared with longitudes of  corresponding profile components
 \citep{i93,gggk02}. Within the conal model, the longitudes of
 profile components are determined by the intersection of the line
 of sight trajectory with the frequency-dependent cones of the
 maximum average intensity, while the longitudes of subpulses are
 determined by the intersection of the line-of-sight trajectory
 with subpulse-associated emission beams, which move
 across the average cones as
 frequency changes \citep{gk96}. We demonstrate in this paper that the
 different frequency dependence of subpulse and profile component
 longitudes is a natural property of the conal model,
  and that both subpulses and
 profile components should demonstrate the same frequency
 dependence of their longitudes within the patchy model.
  We present both general
 qualitative arguments and detailed quantitative model calculations
 to support the above statements. For better understanding of our
 arguments, this paper should be studied along with the paper by
 \citet[][GGGK hereafter]{gggk02}, in which details of frequency dependence of
 emission patterns in PSR B0329$+$54 are discussed within the conal model of pulsar beams.

 \section{Geometry of pulsar radio beams}

 It is widely believed that narrow-band pulsar emission is
 relativistically beamed tangentially to dipolar magnetic field lines.
 Thus, the emission beaming geometry can be described by an
 opening angle $\rho=1^\circ.24~s\ r_6^{1/2}P^{-1/2}$, where 
 $r_6=r(\nu)/R$ is the normalized emission
 altitude (in units of stellar radius $R=10^6$~cm). The mapping
 parameter $0\leq s=d/r_p\leq 1$ is determined by the locus of
 dipolar field lines on the polar cap ($s=0$ at the pole and $s=1$
 at the polar cap edge), where $d$ is the distance from the
 magnetic axis $\bf{m}$ to the field line on the polar cap
 corresponding to a certain detail of the pulse profile (peak of
 subpulse or profile component), and $r_p=1.4\cdot 10^4P^{-1/2}$~cm
 is the canonical polar cap radius. According to the generally - accepted
 concept of the radius-to-frequency mapping, higher frequencies
 are emitted at lower altitudes $r(\nu)$ than lower
 frequencies. \citet{kg97,kg98} found a semi-empirical formula
 describing the altitude of emission region corresponding to a
 given frequency $\nu_{\rm GHz}$ (in GHz) which reads $r_6\approx
 50\cdot\nu_{\rm GHz}^{-0.21}\cdot\tau_6^{-0.1}\cdot P^{0.33}$,
 where $\tau_6$ is the pulsar characteristic age in million years
 and $P$ is the pulsar period. This formula for emission altitudes
 is used in our model calculations.

 To perform geometrical calculations of the radiation pattern one
 has to adopt a model of instantaneous energy distribution on the
 polar cap. Any specific intensity distribution can be transferred
 from the polar cap along dipolar field lines to the emission
 region, and then along straight lines (following the opening angles
 $\rho=1^\circ.24(d/r_p)r_6^{1/2}P^{-1/2}$) to a given observer
 specified by the inclination and impact angles ($\alpha,~\beta$).
 We assume that at any instant the polar cap is populated by a
 number of features with a characteristic dimension ${\cal D}$, 
delivering to the magnetosphere corresponding
 plasma columns flowing along separate bundles of dipolar magnetic
 field lines.
 Each feature can be modelled by the Gaussian intensity distribution.
 Since the elementary pulsar radiation is relativistically beamed along
 the magnetic field lines, we can transform the feature-associated intensity
 pattern to the radio emission region and obtain the subpulse intensity $I$
 observed at the longitude $\varphi$ in the form
 $I_i=\exp(-\kappa l_i^2(\varphi)/{\cal D}^2)$ where the subscript $i$ lebels 
 different features and
 $l_i^2(\varphi)=d^2(\varphi)+d_i^2-2d(\varphi)d_i\cdot
\cos[\sigma(\varphi)-\sigma_{in}]$.
 Each features is located at the position (Fig.~\ref{fig1}) described by the polar 
 co-ordinates $d_i$ (distance from the pole) and $\sigma_{in}=\sigma_{io}+nD_r$
 (magnetic azimuth angle), where the subscript $o$ 
refers to the initial position corresponding to the first pulse,
$n$ is the sequential pulse number and  $D_r$ is the drift rate. 
 The instantaneous emission of the $n$-th pulse is described by
 $I_n(\varphi)=\Sigma_{i=1}I_i(\varphi)$, where  the sum includes a
 number of adjacent features contributing significantly to the  observed 
 intensity, depends strongly on the inclination $\alpha$ and the impact $\beta$
 angles, determining the cut of the line-of-sight tracjectory across the beam 
(Fig.~\ref{fig1}). In fact, the running polar co-ordinates
 along the line-of-sight trajectory can be expressed in the form
 $d(\varphi)=[\rho(\varphi)/1^{\circ}.24]r_pr_6^{-1/2}P^{1/2}$ and
 $\sigma(\varphi)={\rm atan}\left\{\frac{(\sin\varphi\sin\alpha\sin(\alpha+
 \beta)}{\cos(\alpha+\beta)-\cos\alpha\cos\rho(\varphi)} \right\}$,
 and
$\rho(\varphi)=2{\rm asin} 
\left\{\sin^2(\varphi/2)\sin\alpha\sin(\alpha+\beta)+
\sin^2(\beta/2)\right\}$, \citep{ggr84}.
 The average pulse profile is therefore
 $I(\varphi)=\frac{1}{N}\Sigma_{n=1}^{N}I_n(\varphi)$, where $N$ is
 the number of averaged single pulses. 
The geometry of pulsar radiation described
 above \citep[for more details see][GGGK]{gk96,gk97}
 can be applied to both conal and patchy beam models (Fig.~\ref{fig1}).

We adopt a model of a pulsar beam in which the instantenous
 subpulse emission corresponds to a number of isolated subpulse
 beams, while the average emission reflects the conal structure
resulting from circumferential motion of subpulse beams
 \citep{rs75,gk96,gk97,gggk02}. 
 Therefore, the longitudes of subpulse peaks correspond to phases
 of interception of the subpulse beams by the line-of-sight 
 trajectory, while the longitudes of profile component peaks
 are determined by the intersection of the line-of-sight trajectory with
 the average cones.
The important point is that within such  model the subpulse
 enhancements  generally follow  bundles of magnetic field lines 
 different from those of enhancements corresponding to the profile
 components (Fig.~\ref{fig1} - right-hand side). 
 On the contrary, within the
 patchy model both subpulse and profile component enhancements
 follow approximately the same bundles of field lines
 (Fig.~\ref{fig1} - left-hand side). This should be clear
 from the composite Fig.~\ref{fig1}. In fact, subpulse enhancements are
 associated in both models with small subpulse spots $S$ following
 a narrow bundle of dipolar field lines, while the profile
 components correspond to the conical structures $C$ in the conal
 model, and to the narrow patches $PA$ enclosing subpulse spots in
 the patchy model. Because of the diverging nature of dipolar
 field lines controlling the plasma flow, all discussed emission
 features $S$ (spots), $SP$ (sub-patches), $PA$ (patches) and $C$ (cones) are frequency
 dependent, and two frequencies $\nu_1$ and $\nu_2$ are marked for
 each feature in Fig.~\ref{fig1}.

\begin{figure}
\centering
\includegraphics[height=20cm]{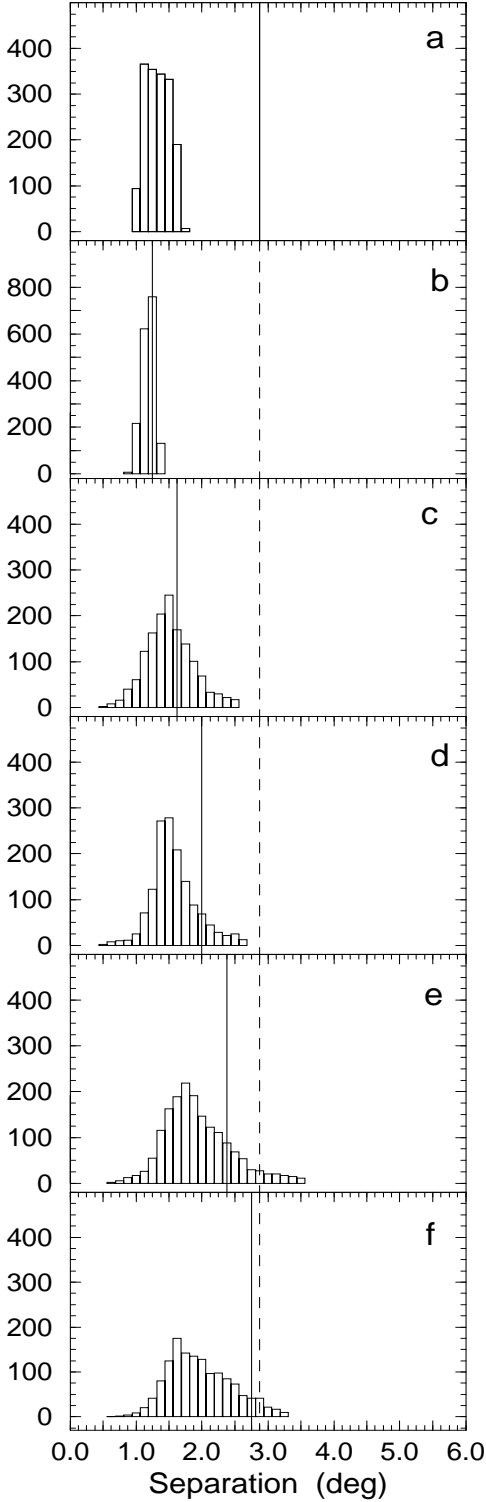}
\caption[]{
Simulation of the distribution of separations $\Delta\varphi_s$
between subpulse peaks as compared with separation $\Delta\varphi_p$ between
the component peaks (vertical solid
lines) measured at two frequencies (400 and 1400 MHz) within
the framework of the conal model (panel a) and various versions of
patchy model (panels b~-~f). The number of pulses (out of the total
number of 2000 simulated) is shown on the vertical axis and the separation
in degrees of longitude is shown on the horizontal axis. The dashed
vertical line going through panels b~-~f is shown to  refer to
the separation $\Delta\varphi_p=2^\circ.875$ in the pure conal model (panel a).
 \label{fig2}}
\end{figure}

\subsection{Conal model of pulsar beams}

The frequency dependence of pulsar emission patterns within the
 angular beaming model of subpulse emission, and the related conal
 model for a mean pulsar beam is illustrated in the
 right-hand side of the composite Fig.~\ref{fig1}. The
 subpulse-associated sub-beams corresponding to HPBW (half
 power beam width) of subpulse emission (thin small circles), which
 are called spots and marked by $S(\nu)$, perform a more or less organized
 circumferential motion around the magnetic axis $\bf{m}$, in which
 the azimuthal angle $\sigma$ varies with time, while the opening
 angle $\rho(\nu)$ remains constant. On average,
this motion determines the
 cones (thick large circles) of the maximum mean intensity with
 the opening angles $\rho_1$ and $\rho_2$ at frequencies $\nu_1$
 and $\nu_2$, respectively. The frequency-dependent longitude
 $\varphi_p(\nu)$ of the profile component is determined by the
 intersection of the line-of-sight  with the average cone
 at the frequency-dependent opening angle $\rho(\nu)$. On the other
 hand, the frequency-dependent longitude $\varphi_s(\nu)$ of
 a subpulse peak is determined by the local maximum intensity along
 the cut of the line-of-sight through the subpulse spot $S(\nu)$.
 When the frequency changes from $\nu_1$ to $\nu_2$, the subpulse
 peak longitude changes from $\varphi_s(\nu_1)$ to
 $\varphi_s(\nu_2)$, while the corresponding profile peak longitude
 changes from $\varphi_p(\nu_1)$ to $\varphi_p(\nu_2)$. Note that
the frequency separation 
$\Delta\varphi_p=\varphi_p(\nu_2)-\varphi_p(\nu_1)$
of profile peaks is generally larger than the frequency separation
$\Delta\varphi_s=\varphi_s(\nu_2)-\varphi_s(\nu_1)$ of subpulse peaks.
 This is consistent with observations published by \citet{i93},
 confirmed recently by GGGK. 
As we argue in the next section, within the
 patchy beam model $\Delta\varphi_p\approx\Delta\varphi_s$,
which is quite different from the conal model in which
 $\Delta\varphi_p\gapp\Delta\varphi_s$.

We  now attempt to quantify our qualitative statements presented
above. We  use the geometrical method of transferring the instantaneous
intensity distribution from the polar cap to the radio emission region 
along the dipolar field lines.
To be able to compare the results of simulations with the observational
data of PSR B0329+54 (GGGK) we adopt the
following parameters $P=0.714$~s, $\dot P=2\cdot 10^{-15}$, $\alpha=20^{\circ}$,
$\beta=4^{\circ}$, $\nu_1=0.4$ GHz and $\nu_2=1.4$ GHz. The conal model is
represented by an arrangement of 12 equisized and equidistant sparks, 
each having the HPBW diameter ${\cal D}\sim 0.2 r_p$, 
circulating
around the magnetic pole at a distance of about 2/3 of the polar cap radius $r_p$.
The results
of simulations are presented in Fig.~2 (panel a). 
The vertical solid line at about $2.9^\circ$
represents the separation $\Delta\varphi_p$ between the peaks of the average
component measured  at the two frequencies. The distribution of
the separations $\Delta\varphi_s$ of subpulse peaks measured at
the two frequencies has a width of about $0.8^\circ$ and peaks around 1.1$^{\circ}$.
The apparent difference between the peak of subpulse separation and
the component peak
separation i.e. $\Delta\varphi_p - \Delta\varphi_s^{\rm peak}\sim 1.8^\circ$,
as well as the width and somewhat skewed shape of the distribution, resemble
the observational data quite well (see Fig.4 in GGGK,
keeping in mind that
zero position bin in their histograms corresponds to cases when
$\Delta\varphi_p=\Delta\varphi_s$).
We have taken into
account only subpulses that corresponded to the same spark at both 
frequencies.
In terms of the observational data this corresponds to subpulses correlating
at two frequencies (refer to the CCF technique used in GGGK).

 \subsection{Patchy model of pulsar beams}

 Within the patchy model of pulsar beams \citep[LM88;][]{m95} the
 observed pulse profile is the product of a ``source function'' and a
 ``window function'', which can be related to the energy/density
 distribution of plasma beams along different bundles of dipolar
 field lines and to properties of the emission mechanism, respectively.
 This model is illustrated schematically on the left-hand side of
 the composite Fig.~\ref{fig1}. The subpulse-associated spots
 marked by $SP(\nu)$ corresponding to the HPBW of subpulse emission
 (thin small circles), occur within the limited areas $PA(\nu)$
 called patches (the simplest and probably unrealistic model is when
 a patch $PA$ can accommodate just one sub-patch $SP$).
The occurrence of $SP(\nu)$ within $PA(\nu)$ can be
 completely random, or more or less organized (including drifting). 
The dependence on frequency $\nu$ reflects the diverging
 nature of dipolar field lines, thus $\nu_2<\nu_1$. The bundle of
 field lines associated with a patch $PA(\nu)$ is only slightly
 larger than each bundle associated with $SP(\nu)$. This
 means that enhancements corresponding to subpulses and profile
 components follow approximately the same bundles of field lines.
 Thus, the frequency dependence of emission patterns in the patchy
 model (left-hand side of Fig.~\ref{fig1}) should be different from
 that of the conal model (right-hand side of Fig.~\ref{fig1}), in which
 subpulse enhancements (spots) and profile components (cones)
 generally follow  quite different bundles of field lines.
If the patch is relatively small as compared to the entire polar cap,
then the frequency separation of the profile
 components $\Delta\varphi_p=\varphi_p(\nu_2)-\varphi_p(\nu_1)$ and
 that of subpulses
 $\Delta\varphi_s=\varphi_s(\nu_2)-\varphi_s(\nu_1)$ should be about the
 same. We demonstrate this below by means of geometrical simulations.

We have calculated a sequence of single pulses and average emission
again for the case of PSR B0329+54, 
assuming that the subpatch $SP(\nu)$ is comparable
in size with the spark associated emission in the conal model, and that the patch $PA(\nu)$
is twice larger than $SP(\nu)$. 
Thus, the projection of a patch onto the polar cap has a characteristic dimension
$P\sim 2{\cal D}\sim 0.4r_p$.
Such a patch encompasses about 30$^{\circ}$ in
magnetic azimuth 
(roughly corresponding to the scale presented in Fig.1 (left-hand side)). The results of
simulations are presented in Fig.~2 (panel b). 
The solid vertical line at about 1$^{\circ}$.25
represents the separation of the profile component associated with $PA(\nu)$ measured
at the two frequencies. The narrow distribution of subpulse peak separations
peaks exactly
at the same value (we have again taken into account only subpulses associated with
the same subpatches $SP(\nu)$ at both frequencies). It is worth noting that if we
chose the sizes of $SP(\nu)$ and $PA(\nu)$ to be equal (which is the simplest and
an unrealistic model of the patchy emission), the distribution would be represented
by a delta function coinciding with the solid vertical line.
The case presented in panel (b) is inconsistent with the frequency dependence of pulsar
radiation patterns (GGGK and reference therein).

We  now start to increase the azimuthal dimension of the patch 
$PA(\nu)$
along the cone, keeping the radial dimension the same (about 2 spark diameters
${\cal D}$ or about $0.4r_p$). Panels c~-~f in Fig. 2 
correspond to the elongation factor
1.3, 1.5, 2.0 and 2.5, respectively. 
Thus, the maximum elongation is equivalent to the length scale comparable with 
the polar cap radius $r_p$.
As one can easily notice from panels c~-~f, increasing elongation
results in two effects: (1) the separation of the component peaks increases towards
the value corresponding to the conal model (dashed vertical line)
i.e $\Delta\varphi_p \to 3^\circ$, while
the distribution of subpulse peak separations $\Delta\varphi_s$ gets broader and
broader and peaks at correspondingly larger and larger
distances from $\Delta\varphi_p$. Moreover, the skewed shape of the distribution
becomes more and more apparent in panels (d) and (e), to such an extent that
the case presented in  panel (f) seems almost
undistinguishable from the pure conal case presented in  panel (a),
except that the width of the distribution is too large compared with observations
(see Fig. 4 in GGGK).
Further increasing
of the elongation factor beyond 2.5 (corresponding to 60$^\circ$ of magnetic azimuth or
1/6 of the full cone) does not practically change the results of the simulations.
This means that the unrealistic
patches  elongated to a large extent along circles centered on the
magnetic axis would resemble the conal model. Although this conclusion seems trivial,
it allows us to constrain some characteristics of both patchy and conal models.
This is described in the two following paragraphs.

First we can ask: what is the probability that an adequately large 
and favourable patch elongated along a cone and 
corresponding to panel (f) mimics
the conal model presented in panel (a) of Fig.~2. 
Of course, we have to take into account
that another similar patch is required on the opposite side of the fiducial
plane (see Fig.1) to account for the second outermost component of PSR B0329+54.
Let us assume for simplicity that our elongated patch is a rectangular figure
with the shorter side $A\approx 0.4r_p$ and the longer side $B\sim 1.0r_p$
(see above for an estimate of the dimensions). Thus, the surface area of such a patch 
$S_{patch}\approx A\cdot B=0.4r_p^2$
and the probability of its occurence  in any location of 
the polar cap is ${\cal P}_1=S_{patch}/S_{cap}=0.4r_p^2/(\pi r_p)^2\approx 0.13$.
The probability of occurence of another such patch somewhere on the remaining part
 of the polar cap is ${\cal P}_2=S_{patch}/(S_{cap}-S_{patch})\approx 0.15$.
To estimate the probability of the proper alignment along the cone, we can
calculate a number of different independent orientations of our 
rectangular figure inscribed into a circle of the diameter approximately 
equal to $A=r_p$. 
One can easily show that the number of independent orientations
is approximaly $2\pi(r_p/2)/(0.4r_p)\sim 8$ and thus the probability of
alignment along a cone ${\cal P}_0\sim 0.12$. So far, the resultant probability is
${\cal P}={\cal P}_1\cdot {\cal P}_2\cdot {\cal P}_0^2\approx 3\cdot 10^{-4}$.
The requirement of having two such patches symmetricaly placed with respect to
the fiducial plane 
only decreases the final probability. Thus, we can conclude that
mimicking the conal radiation pattern by a specially arranged patchy distribution is 
extremly unlikely, at least in the case of PSR B0329+54 and other pulsars showing 
similar frequency dependence of emission patterns (GGGK and references
therein).

We can now approach the results of our elongation exercise from a different angle.
Since further elongation beyond the case presented in panel (f) does not
change the obtained distribution, we can constrain the possible spread of the cone
in the radial direction
as compared with the ideal (and probably not realistic) case presented in panel (a).
Remembering that the initial patch size $PA$
(panel b) was twice $SP$, with $SP$ corresponding to the spark size $\cal D$,
we conclude that the realistic version of
the cone can accommodate up to two sparks in radial dimension.
This means that a locus of maximum intensity within the average pulsar beam has the form
of a narrow ring rather than a circle.

\begin{figure*}
\centering
\includegraphics[width=17cm]{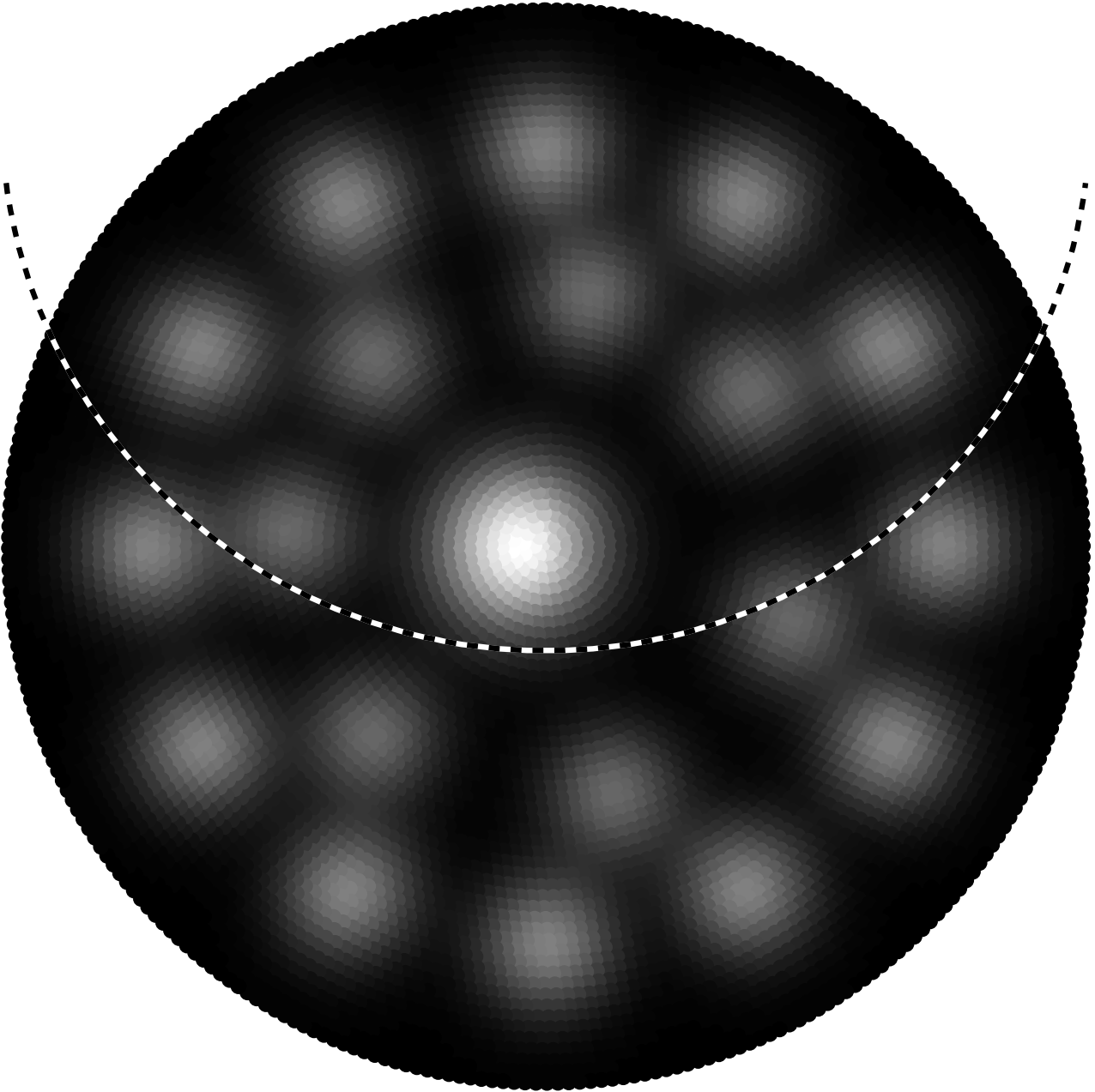}
\caption[]{The proposed geometry of observation and
 the instantenous arrangement of sparks on the polar cap for
 conal model of pulsar beam of PSR J1834$-$0426. Notice that about 30\%
of the line-of-sight (l-of-s) 
trajectory covers the beam, resulting in very broad ($\sim 130^{\circ}$)
profile of this pulsar.
 \label{fig3}}
\end{figure*}

 \subsection{Beam reconstruction techniques:}

 Recently Han \& Manchester (2001; HM01 hereafter) attempted to
 reveal the shape of pulsar radio beams. They claim to have constructed a
 two-dimensional image of the ``average'' mean pulsar beam using
 a special technique applied to all available multi-component pulse
 profiles with good quality polarimetric data. They mapped the
 observed profile intensity onto the line-of-sight crossing the
 normalized polar cap at the normalized impact angle $\beta_n$
 estimated from the polarization angle swing. To include a second
 dimension (perpendicular to the line-of-sight), they broadened the
 distribution in latitude applying a Gaussian to each
 longitudinal sample. Adding all pulsars together they obtained a global
 average beam pattern, which (after normalization to correct for the nonuniform
 distribution of $\beta_n$) they believe represents the global mean pulsar
 beam. HM01 concluded that their results are consistent with the patchy
 rather than conal beam model
 (see their Fig.~4). However, as we demonstrate in
 items (i)-(iv) below, their beam reconstruction technique is not
 general enough to reveal the true structure of pulsar beams.

\begin{figure*}
\centering
\includegraphics[width=17cm]{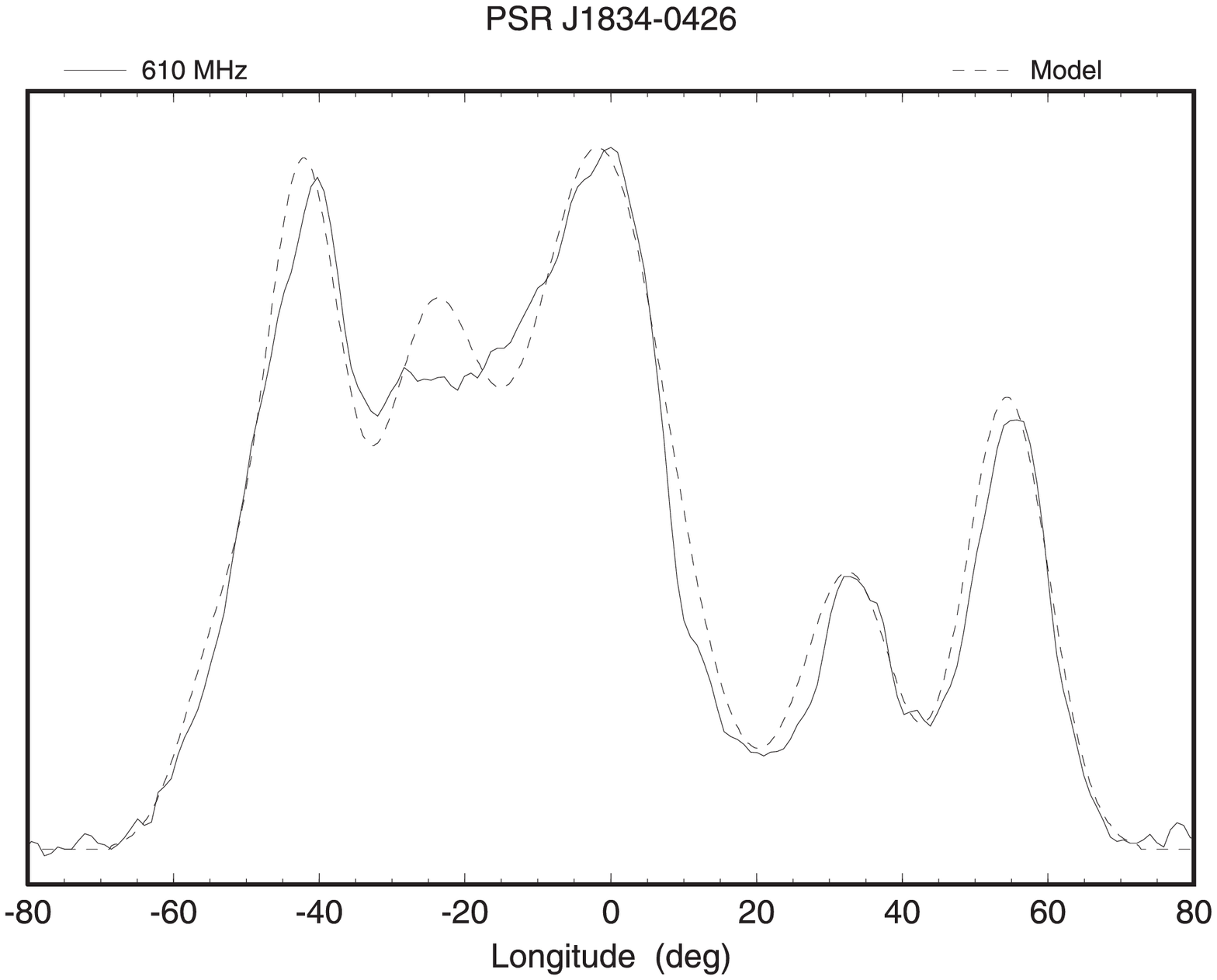}
 \caption{The observed (solid line) and simulated
   (dashed line) mean profile of PSR J1834$-$0426. The conal
  model calculations corresponding to 610 MHz were performed
  following the instantaneous energy distribution (presented
 in Fig.~\ref{fig3}). The geometry of observations was
  determined by the inclination angle $\alpha=10^\circ$ and
   the impact angle $\beta=2^\circ.5$. The adopted emission altitude
   $r(610~{\rm MHz})\approx 30R$. The 610 MHz observational data 
   \citep{gl98} were taken
   from the European Pulsar Network data base
   (http://www.mpifr-bonn.mpg.de/div/pulsar/data/)
\label{fig4}}
\end{figure*}

 \begin{description}
 \item (i) HM01 projected all emission features onto a normalized polar
 cap, ignoring the dependence of the beaming angles on the emission
 altitude, which most probably depends both on the radio frequency and
 on the pulsar period and its derivative 
\citep{kg97,kg98,k01}. Taking into account the
 diverging nature of dipolar field lines with increasing altitudes,
 the projection of the emission pattern onto the polar cap must be
 performed more carefully. HM01 used  data at frequencies between
 600 MHz and 1.6 GHz, which probably takes care of frequency dependence of
 the opening angles $\rho(\nu)\propto
 r^{1/2}(\nu)\propto(\nu)^{-0.1}$. However, the period
 dependence $\rho(p)\propto P^{\sim 0.2}$ is stronger and can
 result in a broader spread of projections onto the normalized polar
 cap for short and long period pulsars, respectively. 

 \item (ii) A similar
 problem concerns the putative conal structure of pulsar beams which
 HM01 did not reject a priori. However, they
 assumed that adding
 the normalized polar caps of different pulsars would not smear the
 conal structure, if it existed. It seems that the conal patterns are
 almost certainly different in different pulsars. MD99
 showed that the pulsar emission beams follow a nested cone
 structure with up to three distinct cones, although only one or
 more of the cones may be active in a given pulsar \citep[see
 also][]{r93,gks93}. Also GS00 argued that the number of cones is
 a function of the basic pulsar parameters $P$ and $\dot{P}$. 

 \item (iii) The analysis of HM01 is based on the orthogonal
normalized impact parameter $\beta_n=\beta_{90}/\rho_{90}$ ,
where $\beta_{90}$ and $\rho_{90}$ are the impact angle $\beta$ and the
beam radius $\rho$ computed for the inclination angle $\alpha=90^{\circ}$
(LM88). 
The orthogonal normalized impact parameter $\beta_n$ differs
from the actual normalized impact parameter $\beta/\rho$ in many aspects
\citep[e.g.][]{gks93}. We would like to emphasize here that $\beta/\rho$
depends on the pulsar period $P$, the inclination angle $\alpha$ and
the observing frequency $\nu$. These dependences, which can affect the emission
patterns, are not accounted for in the HM01 analysis based on the orthogonal
impact parameter. 

\item(iv) HM01 included the core components in their analysis. This is
another possible source of confusion, since the core components are rather
randomly placed with respect to the midpoint of the overall profile.
For this reason MD99 excluded the core components from their analysis,
which revealed a nested cone structure of pulsar beams.
\end{description}

Given the problems listed above, we conclude that the results of the HM01's
beam reconstruction are illusive. The lack of an apparent conal structure in their
'global' pulsar beam does not exclude the existence of such structures in
particular pulsars, expecially if they are determined by physical and geometrical
factors, which may vary a great deal among different pulsars. HM01 demonstrated
that the conal emission is not confined to a single region at the beam
boundary. They also concluded that if multiple cones exist, they are at 
different radii relative to the beam radius in different pulsars. 
This is exactly what the multi-nested
cone model of GS00 predicts. Both the number of cones and the relative radius of
a given cone depend on a pulsar period and its derivative in their model.

As an example of pulsar modelling within the multi-nested cone scenario
(GS00), we use the case of PSR J1834$-$0426. The patchy model for this pulsar
was presented by HM01 (their Fig.~1). This pulsar has a very broad profile with 
the pulse width $W\sim 130$ degrees of longitude, which implies a small inclination
angle $\alpha\ll 90^{\circ}$. Assuming a realistic emission altitude 
$r_6\sim 40$ \citep{kg97,kg98}
we estimated the observational angles $\alpha\sim 10^{\circ}$
and $\beta\sim 2^{\circ}.5$ (consistent with $\alpha=7.9^{\circ}$ and 
$\beta=1.6^{\circ}$ given by LM88). Figure 3 shows an instantenous arrangement of
sparks on the polar cap of PSR J1834$-$0426, obtained from the complexity parameter
of GS00. The circumferential motion of these sparks
results in the average structure of two nested cones (e.g. Fig.~6 in GGGK).
Notice that about 35\% of the line-of-sight trajectory stays within the beam, which
leads to a broad $\sim 130^{\circ}$ average profile presented in Fig.~4.

 \section{Discussion and conclusions}

 In this paper we explore a geometrical method of pulsar radiation 
simulation, based on two  well-justified assumptions:
 (i) the elementary coherent radio emission is narrow-band, and the
 emission altitude depends on both the frequency and the pulsar
 period,
(ii) the emission is relativistically beamed tangently to
 dipolar magnetic field lines.
We have considered two competitive models of the organization of pulsar emission beams:
 the conal model, in which enhancements related to subpulse emission in single pulses
are distributed along the cones 
corresponding to maximum average intensity, 
and the patchy beam model in which subpulse enhancements corresponding to the
component of the mean profile are confined to the patchy area limited both in azimuthal
and radial dimensions. 
We examined the consistency of these rival models with the variety of observational data.

 We have argued that a number of observational
 properties of pulsar radio emission, namely: (i) binomial
 distribution of the opening angles \citep{r93,gks93,kwj94}; (ii)
 high impact angles corresponding to single and double profile
 pulsars (MD), and (iii) different frequency dependence of
a subpulse and corresponding profile component longitudes
 \citep{i93,gk96,gggk02}, strongly support the conal model of
 pulsar beams. The alternative patchy beam model is inconsistent
 with these observational properties of pulsar radiation.

We have
also demonstrated that the beam reconstruction technique developed
 by \citet{hm01} is not capable of  revealing the true structure of
pulsar beams.
In fact, their formalism assumes implicitely
that neither the radio emission
 altitude nor the number of putative nested-cones and their
 locations within the pulsar depends on the pulsar period. 
The lack of an apparent conal structure in their ''global beam'' does not
exclude the conal beam model.
Thus, the results of the HM01 analysis provide no
 strong evidence of  patchy beam structure in pulsars.

 We tend to favour the version of the conal model in which the
 relationship between the subpulse-associated beams and cones of
 the average emission is established through the phenomenon of the
 ${\bf E}\times{\bf B}$ drift \citep[][GS00]{rs75,dr99,dr01}, which
 forces the spark filaments of plasma to rotate slowly around
 the magnetic axis. This ``spark model'' of radio pulsars was
 recently tested statistically by \citet{fan01}. They showed,
 by means of  Monte Carlo simulations, that various pulsar
 parameters can be reproduced if both the spark dimension and
 their mutual separation are approximately equal to the height $h$
 of the polar gap \citep[][GS00]{rs75}, 
or consequently, the maximum number of sparks
 along the diameter of the polar cap with the radius $r_p$ is
 $N_{max}\sim r_p/h$. Thus, their conclusions are consistent with
 the assumptions used in this paper.

We note that  evidence of a relationship
between drifting subpulses and the conal structure of mean pulsar
beams already exists  in the literature. \citet{hw87} examined
three pulsars with triple average profiles showing subpulses
drifting across the full pulse window (including the central
component). They found that these pulsars are consistent with two
nested cones of emission, each associated with a prominent
subpulse drift. Moreover, \citet{pw86} showed that in complex
profile pulsars there is a strong correlation between drifting
subpulses associated with different profile components. Such
correlations are natural within the ${\bf E}\times{\bf B}$ induced
conal model, but inconsistent with the patchy model of pulsar beams.

Finally, we suggest that more pulsars should be observed in the single pulse,
simultanenous dual frequency mode. Our simulation method illustrated
and described in Fig. 2 can be adapted to the analysis of such
data in order to ultimately discriminate  between conal and patchy beam models
in pulsars.

 \begin{acknowledgements}
This paper is supported in part by the Grant 2~P03D~008~19 of the
Polish State Committee for Scientific Research. We are grateful to
the referee
Dr. D. Mitra for extremely helpful comments that improved the final 
paper. We also thank Prof. Dr. R. Wielebinski for hospitality
at the MPIfR in Bonn, where this work was completed.
We thank E. Gil and M. Wujci\'ow for technical assistance.
\end{acknowledgements}

\end{document}